\documentclass[11pt]{article}
\usepackage{moriond,epsfig}

\def\Journal#1#2#3#4{{#1} {\bf #2}, #3 (#4)}

\def\aap{A\&A}
\def\apj{ApJ}
\def\apjl{ApJL}
\def\mnras{MNRAS}
\def\aj{AJ}
\def\nat{Nature}

\def\be{\begin{equation}}
\def\ee{\end{equation}}
\def\bea{\begin{eqnarray}}
\def\eea{\end{eqnarray}}

\begin{document}
\vspace*{4cm}
\title{Relativistic Particles in Clusters of Galaxies}

\author{ T. A. En{\ss}lin }

\address{Max-Planck-Institut f\"{u}r
Astrophysik, Karl-Schwarzschild-Str.1, Postfach 1317, D-85741 Garching,
Germany}

\maketitle
\abstracts{
A brief overview on the theory and observations of relativistic
particle populations in clusters of galaxies is given. The following
topics are addressed: (i) the diffuse relativistic electron population
within the intra-cluster medium (ICM) as seen in the cluster wide
radio halos and possibly also seen in the high energy X-ray and
extreme ultraviolet excess emissions of some clusters, (ii) the
observed confined relativistic electrons within fresh and old radio
plasma and their connection to cluster radio relics at cluster merger
shock waves, (iii) the relativistic proton population within the ICM,
and its observable consequences (if it exists), and (iv) the confined
relativistic proton population (if it exists) within radio plasma. The
importance of upcoming, sensitive gamma-ray telescopes for this
research area is highlighted.}

\section{Introduction}

\begin{figure}
\begin{center}
\psfig{figure=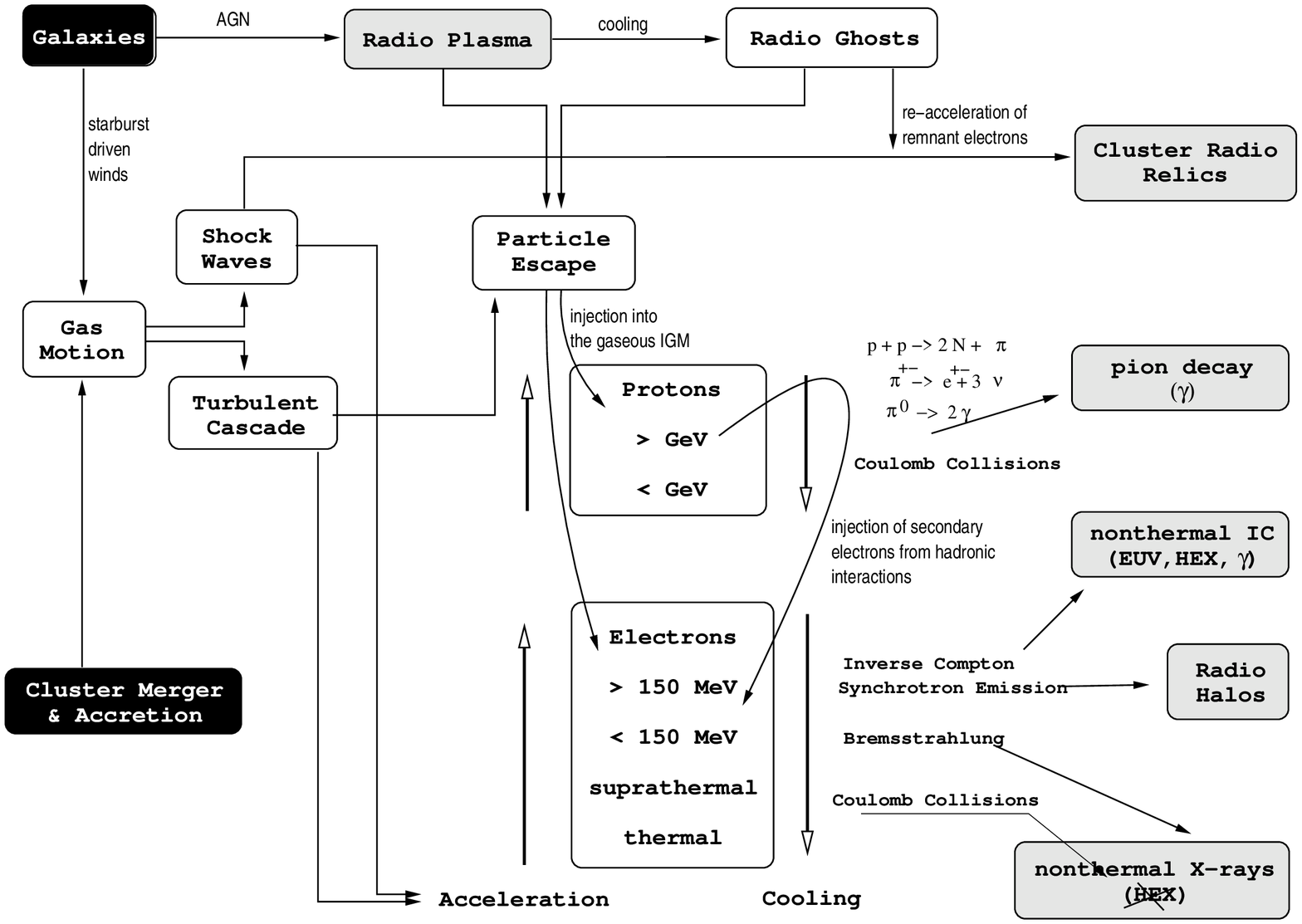,width=0.85\textwidth}
\end{center}
\caption{\label{fig:diag} Basic theory building blocks covering most of
the proposed scenarios for the production and maintenance of
relativistic particle populations in galaxy clusters and their
observational consequences.  For details see text.}
\end{figure}

Even though the study of relativistic particle population is more than
three decades old, it has recently received a significant increase in
attention by various researchers. Here, a brief and, therefore,
incomplete and personally biased overview of this field is
provided. A guide through the lines of argumentation is given by
Fig. \ref{fig:diag}, which sketches the main dependencies of the
components of the theory and their observational consequences. This
figure is explained in the following.

The main energy sources of the relativistic particle population
in clusters are outflows from galaxies (galactic winds, radio
jets) and/or the energy released in accretion on galaxy clusters. The
first sources can directly eject relativistic particles into the
ICM \cite{1977ApJ...212....1J,1993ApJ...406..399G}, whereas the latter
produce shock waves and turbulence, which can accelerate particles via
the Fermi
mechanisms \cite{1987A&A...182...21S,1997A&A...321...55D,1997MNRAS.286..257K,1998AA...332..395E,2001MNRAS.320..365B,2001ApJ...559...59M,2001ApJ...562..233M}. Also
the termination shocks of galaxy winds were proposed as shock
acceleration sites \cite{1996SSRv...75..279V}. A different source of
relativistic particles may be the annihilation of certain dark matter
particles \cite{2001ApJ...562...24C}.

The relativistic particles loose energy via various radiative and
non-radiative processes, allowing to measure or constrain their
spectral energy distribution observationally. The most important
loss channels are discussed in the following.

\section{Relativistic Electrons}
\subsection{Diffuse Population: Radio Halos}
The existence of relativistic electron populations in
galaxy clusters is known since Willson's \cite{1970MNRAS.151....1W}
(1970) detection of extended diffuse radio emission from the
intra-cluster medium of Coma. Today, there exists a much larger sample
of similar
detections \cite{1996IAUS..175..333F,1999dtrp.conf....3F,Feretti.Pune99,2000NewA....5..335G,2001ApJ...548..639K}.
This radio emission is believed to be synchrotron emission from highly
relativistic ($\gamma \sim 10^4$) electrons spiraling in intra-cluster
magnetic fields of $\sim\mu$G strength. The origin of the fields and
particles were unclear at the time of the first detection, and are
still a puzzle, although substantial progress has been made in their
understanding. A clue is the obvious correlation of the presence of
extended radio emission with the presence of cluster substructure as an
indication for ongoing or recent cluster merger
activity \cite{2001ApJ...553L..15B}. This indicates that the main
energy source of the electron population are merger shock waves and
possibly merger induced turbulence \cite{2001ApJ...563..660F}.

Extended cluster radio emission, which is not associated with
individual galaxies, is nowadays classified as {\it cluster radio
halos} and {\it cluster radio relics}. Cluster radio halos are steep
spectrum radio sources, which often have morphologies very similar to
the X-ray emission of the cluster \cite{2001A&A...369..441G} indicating
that there is a direct link between their energetics and that of the
ICM. It is therefore likely that the emitting electrons occupy the
same subvolume as the thermal X-ray emitting ICM gas, which we assume
in the following. We denote this population as the {\it diffuse
population of relativistic electrons} in galaxy clusters. It is worth
noting, that the cluster wide radio halos cannot be the direct result
of passing merger shock waves, since the cluster crossing time of a
shock wave is much larger ($\sim 10^9$ years) compared to the radio
emitting electron cooling time ($\sim 10^8$ years). Since there is
some indication that radio halos only appear in regions which were
passed by a merger shock
wave \cite{2001ApJ...559..785K,2001ApJ...563...95M} one can speculate
if some agent stores some fraction of the shock released energy and
provides it successively to the radio electron population. Possible
natures of such an agent are plasma
turbulence \cite{2001MNRAS.320..365B} or a shock accelerated population
of relativistic protons \cite{1980ApJ...239L..93D} (see below).

\subsection{Confined Populations: Radio Cocoons, Radio Ghosts, \&
Cluster Radio Relics} 

In contrast to the radio halo electrons spatially {\it confined populations of
relativistic electrons} exist in galaxy clusters: There are the
electrons released by outflows from radio galaxies and confined from
the thermal ICM by strong magnetic fields within the so called {\it
radio cocoons}. Since the higher energy electrons rapidly lose energy
by synchrotron and inverse Compton (IC) emission, the observable radio
emission of an old radio cocoon disappears after $\sim 10^8$
years. Afterwards, the relativistic electrons (and any other particle
population) should still be confined by the magnetic fields. Recent
support for the existence such of non-emitting {\it radio
ghosts} \cite{Ringberg99} is given by the rapidly growing number of
detections of cavities  \cite{2002A&A...384L..27E,2002astro.ph..1325S}
in the X-ray emitting gas of clusters of galaxies, sometimes filled
with observable radio emission, sometimes not.

\begin{figure}
\begin{center}
\psfig{figure=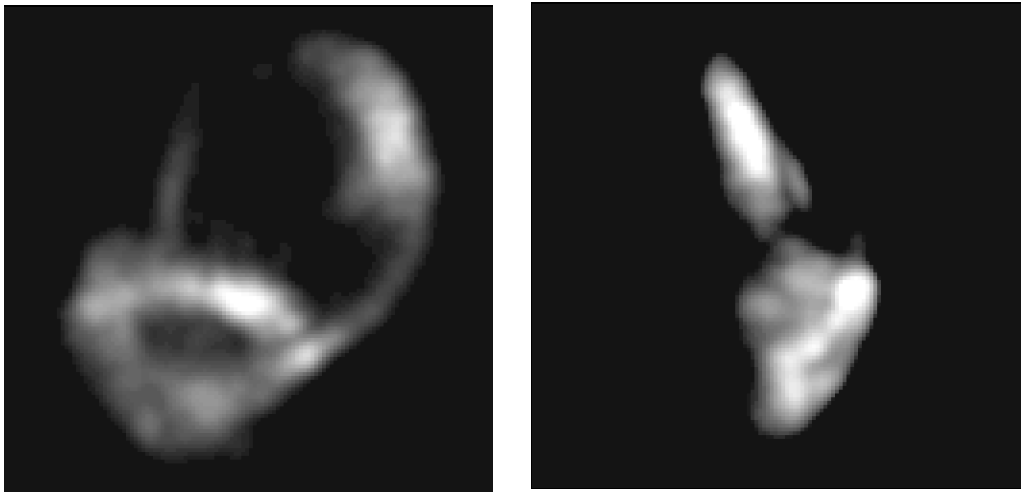,width=0.85\textwidth}
\end{center}
\caption{\label{fig:relic} Simulated cluster radio relic: radio
emission of a radio ghost after shock passage. Left: face on view on
the shock plane, right: edge on view on the same relic.}
\end{figure}
 
Furthermore, another class of objects with possibly confined electron
populations are the cluster radio relics. They are typically
located at the periphery of clusters, sharply edged, and
sometimes of a filamentary morphology. Their steep spectrum radio
emission is often highly polarized, which highlights their physical
distinctness from the unpolarized radio halos.

There are now several examples of a spatial co-location of relics and
cluster merger shock
waves \cite{1998AA...332..395E,2001ApJ...559..785K,1998ApJ...500..138D,1999ApJ...518..603R,1998ApJ...493...62R}
indicating that the electrons observed there in radio were directly
accelerated by the shock. Two shock acceleration mechanisms were
proposed so far for relics: direct diffusive shock acceleration out of
the thermal
pool \cite{1998AA...332..395E,2001ApJ...562..233M,1999ApJ...518..603R},
and adiabatic compression of an old electron population in a radio
ghosts \cite{2001A&A...366...26E,2002MNRAS.331.1011E}. Both processes
may be realized in nature: E.g. the giant radio relics in Abell 3667
seem to be of the first
type \cite{1998AA...332..395E,1999ApJ...518..603R,1997MNRAS.290..577R},
whereas the small, filamentary relics in Abell 85 are probably of the
second type \cite{2002MNRAS.331.1011E,1998MNRAS.297L..86S} (see
Fig. \ref{fig:relic}).

\subsection{Observational Constraints on the Spectrum}

\begin{figure}
\begin{center}
\psfig{figure=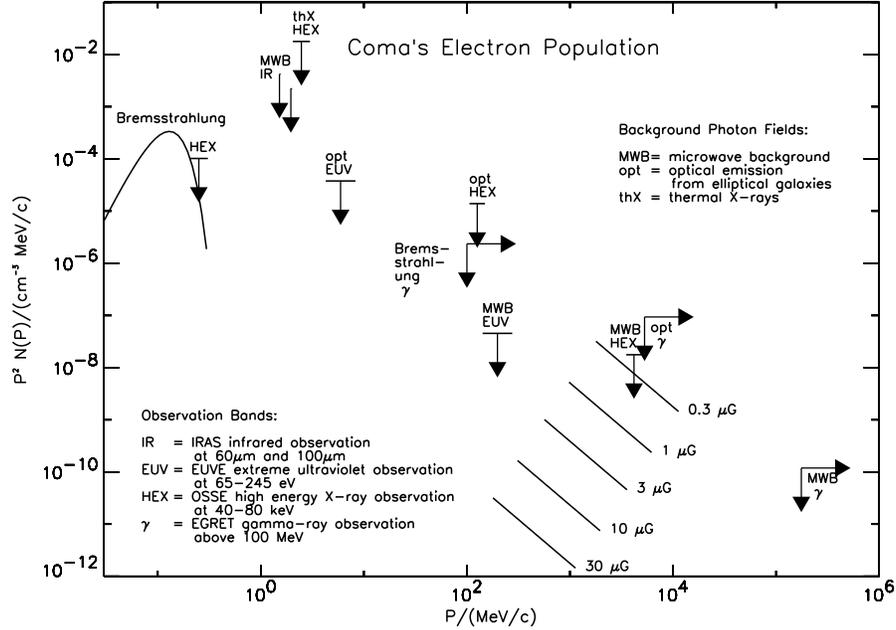,width=0.8\textwidth}
\end{center}
\caption{\label{fig:e} The central electron spectrum in the Coma
cluster of galaxies.   For details see text.}
\end{figure}

Relativistic electrons should reveal their presence via a variety of
emission processes. They can up-scatter a present photon population
to higher energies. They can produce non-thermal bremsstrahlung, and
they emit synchrotron radiation if located in magnetic fields.

Fig. \ref{fig:e} is a compilation of the available observational
information on the electron spectrum in
{Coma \cite{1998AA...330...90E}}$^{\!,\,\,}$\footnote{The compilation was
done in 1997, but since then only the upper limit on the HEX excess by
OSSE was accomplished by a detection of a only slightly lower flux by
Beppo-Sax in 1998. Therefore the figure can still be regarded to be up to
date.}.  The line on the left hand side is the thermal distribution.
The line on right hand side gives the required electron populations in
order to produce the observed radio halo of Coma with the labeled
magnetic field strengths.  The given upper limits result from
observational upper limits to bremsstrahlung or IC scattering fluxes
in various wave-bands. Two of them correspond to actual detections:
the extreme ultraviolet (EUV) \cite{1996Sci...274.1335L} and the high
energy X-ray (HEX) \cite{1999ApJ...513L..21F} excess fluxes. Some of
the corresponding upper limits (labeled by EUV and HEX) on the
electron spectra could therefore be data points. For further details
of this figure see En{\ss}lin \& Biermann \cite{1998AA...330...90E}.

The explanation of the HEX excess is problematic. An IC interpretation
in terms of up-scattered CMB photons requires magnetic field strength
of $\sim 0.1\mu$G in order to be consistent with the observed radio
halo flux level
\cite{1979ApJ...227..364R,1994ApJ...429..554R,1999ApJ...513L..21F,1999ApJ...520..529S}.
This is much lower than the $\sim 1\mu$G fields suggested by Faraday
rotation measurements \cite{1994RPPh...57..325K,2001ApJ...547L.111C,2001astro.ph.10655C}.
There might be ways to reconcile these measurements, e.g. by
inhomogeneously and anti-correlated relativistic electron and magnetic
field distributions \cite{1999AA...344..409E}, but they seem to be a
little contrived. Another proposal, that the HEX excess is actually
due to a supra-thermal electron population producing bremsstrahlung
\cite{1999AA...344..409E}, is even less likely due to the inefficiency
of bremsstrahlung compared to the unavoidable huge Coulomb losses
\cite{2001ApJ...557..560P}.

It is worth noting, that gamma-ray observations of up-scattered optical
photons will probe exactly the same energy range as probed by HEX
observations of up-scattered CMB photons. The sensitivity of upcoming
gamma-ray telescopes should therefore allow to give important insight
into this puzzle. In addition to this, such gamma-ray observation will
also probe for the presence of TeV
electrons \cite{2000Natur.405..156L,2002astro.ph..3014M,2002astro.ph..2318K}.

\section{Relativistic Protons}

\subsection{Diffuse Populations: Origin of Radio Halos?}

Today, a direct proof of the presence of a relativistic proton
populations in galaxy clusters is still lacking although our
knowledge of the galactic cosmic rays suggests the presence of a much
more energetic proton population compared to the electrons. Protons
are long-lived in the ICM, with lifetimes of the order of a Hubble
time. Since spatial diffusion of moderate energy protons (say below
$10^{15}$ eV) is slow, even on a cosmological timescale, they should
basically be stored within the
cluster \cite{1996SSRv...75..279V,1997ApJ...477..560E,1997ApJ...487..529B}.

A diffuse relativistic proton population can, in principle, be observed
by secondary particle produced in hadronic interactions with the
background gas nucleons. The hadronic production of neutral pions
leads to gamma-ray emission from galaxy
clusters \cite{1982AJ.....87.1266V}, which may be detectable with
upcoming gamma-ray telescopes. The non-detection of the Coma cluster
by the EGRET telescope limits the energy density of protons to be
below the thermal electron density of the ICM at least in the case of
Coma \cite{1996SSRv...75..279V,1997ApJ...477..560E,1998APh.....9..227C,1999ApJ...525..603B,2000A&A...362..151D}.
Upcoming gamma-ray instruments will either detect such emission, or
further constrain the relativistic proton population.

The hadronic production of charged mesons leads to the injection of
electrons and
positrons \cite{1980ApJ...239L..93D,1982AJ.....87.1266V,2000A&A...362..151D,1999APh....12..169B}
into the ICM and to the emission of
neutrinos \cite{2000A&A...362..151D}. The neutrinos are practically
undetectable with present and planned telescopes, but the injected
electrons could be the ones seen in radio halos.

\subsection{Confined Populations: Undetectable?}

A population of relativistic protons within radio plasma if it exists,
is practically unobservable by direct means. However, improvement in
the understanding of radio jets, or detailed analysis of the
mechanical properties of radio plasma \cite{1992A&A...265....9F} may
lead to firmer conclusions in the future.

\section*{Acknowledgments}

It is a pleasure to thank Berit Uhlmann and Marcus Br{\"u}ggen for
helping me writing this article.

\section*{References}

\bibliographystyle{unsrt}


\end{document}